\documentclass[12pt]{iopart}
\usepackage{epsfig}
\begin{document}

\title[Full counting statistics of a chaotic cavity]
{Full counting statistics of a chaotic cavity with asymmetric leads}

\author{Oleg M. Bulashenko}
\address{Departament de F\'{\i}sica Fonamental,
Universitat de Barcelona, Diagonal 647, E-08028 Barcelona, Spain}

\ead{oleg.bulashenko@ub.edu}

\begin{abstract}
We study the statistics of charge transport in a chaotic cavity attached to
external reservoirs by two openings of different size which transmit non-equal 
number of quantum channels.
An exact formula for the cumulant generating function has been derived 
by means of the Keldysh-Green function technique within the circuit theory
of mesoscopic transport. 
The derived formula determines the full counting statistics of charge
transport, i.e., the probability distribution and all-order cumulants of 
current noise.
It is found that, for asymmetric cavities, in contrast to other mesoscopic 
systems, the third-order cumulant changes the sign at high biases.
This effect is attributed to the skewness of the distribution of transmission 
eigenvalues with respect to forward/backward scattering.
For a symmetric cavity we find that the third cumulant approaches 
a voltage-independent constant proportional to the temperature and the number 
of quantum channels in the leads.
\end{abstract}


\maketitle

\eqnobysec

\section{Introduction}

Statistical properties of charge transfer in mesoscopic conductors 
have recently received rapidly growing attention. 
A powerful theoretical approach to the fluctuation problem is based on 
the concept of full counting statistics (FCS) \cite{LLL}.
It gives the probability to transmit a given number of charges from one
terminal of a mesoscopic conductor to another terminal during the observation 
time.
{}From the knowledge of these probabilities one can readily derive 
not only average current and noise, but all higher cumulants of current
fluctuations and extreme-value statistics. 
Thus FCS represents the complete information which can be gained 
from charge counting in a transport process.

Since the introduction of FCS for electron counting in mesoscopic conductors 
by Levitov and co-workers \cite{LLL},
the approach has been successfully applied to various systems including 
quantum point contacts, diffusive conductors, cavities, metal and 
superconductor junctions \cite{nazarov-book}.
Until very recently, the experimental measurements were focused mainly on 
shot noise---the second cumulant of the fluctuating current
\cite{blanter-buttiker00}.
Experimental investigation of FCS has been opened up by measurements of 
the third cumulant of current fluctuations \cite{reulet03,bomze05}.
Since the third cumulant (odd-order cumulant) changes sign under a time
reversal transformation, it is not masked by thermal fluctuations and can be
observed at relatively high temperatures \cite{levitov01}.
In contrast, accurate measurements of the second cumulant or shot noise 
(even-order cumulant) is difficult because of the requirement to maintain 
a low temperature at a high applied voltage.

In this paper we address the FCS of an asymmetric chaotic cavity 
with two attached leads supporting non-equal number of quantum channels.
Previous studies on asymmetric chaotic cavities include the following.
The second cumulant has been calculated by Nazarov \cite{nazarov95} 
by means of a circuit theory, by Beenakker \cite{beenakker97} using 
random-matrix theory (RMT), by Blanter and Sukhorukov semiclassically
\cite{blanter00}, and by Schanz \etal \cite{schanz03} 
using quantum graphs. The third and fourth cumulants of current were obtained 
by Blanter \etal \cite{blanter01}.
The authors found a discrepancy between semiclassical
and RMT results for high-order cumulants starting from the fourth cumulant.
Subsequently, Nagaev \etal \cite{nagaev02-2} were able to explain the
discrepancy by taking into account ``cascade correlations'' \cite{nagaev02-1} 
between the cumulants within a semiclassical approach.
It was pointed out, that evaluation of high-order cumulants goes beyond
the validity of the standard Boltzmann-Langevin scheme: one has to use
cascade diagrams, or to introduce non-local noise correlators in the 
Boltzmann-Langevin equation \cite{gutman-book}.
It should be noted that the results on the third and forth cumulants 
of current fluctuations \cite{blanter01,nagaev02-2}
were obtained in the zero temperature limit only.
Experimental measurements of the second cumulant (shot-noise) on chaotic 
cavities were performed by Oberholzer \etal 
\cite{oberholzer01,oberholzer02}.
Recently, a stochastic path integral approach has been formulated by
Pilgram \etal \cite{pilgram02} 
and Jordan \etal \cite{jordan04}, which is efficient to study FCS
for arbitrary mesoscopic semiclassical systems including stochastic networks.
The authors \cite{jordan04} have derived the diagrammatic rules to calculate 
the high-order cumulants of current fluctuations which take into account 
the Nagaev's cascade correction terms \cite{nagaev02-1}.
This approach was applied to calculate the cumulants of charge fluctuations 
in asymmetric cavities in the regime of strong electron-electron scattering 
\cite{pilgram02,pilgram04}.
Pilgram and B\"uttiker studied the effect of screening on fluctuations of 
charge inside the cavity \cite{pilgram03}.

In this paper, we use the extended Keldysh-Green function technique within
the circuit theory introduced by Nazarov \cite{nazarov99-1}.
By using this technique, we have obtained the cumulant generating function 
(CGF) in exact analytical form which contains complete information about
all-order cumulants of current fluctuations in asymmetric chaotic cavities 
at arbitrary bias-to-temperature ratio.
The analytical formula for CGF is expressed in terms of the anticommutator 
of the matrix Green functions in the leads. 
In this form it can be applied to calculate all-order cumulants of current 
noise and extreme-value statistics for different types of the contacts.
In section \ref{sec-CGF}, we present the derivation of CGF for asymmetric 
cavities and show the correspondence with the results obtained earlier 
in the symmetric limit. The distribution function of charge transfer 
is obtained in section \ref{sec-count}.
The first three cumulants of current fluctuations are analyzed
in section \ref{sec-cum} for arbitrary bias-to-temperature ratio. 
The third-order cumulant was found to be linear in voltage at both low
and high biases, but, in contrast to other mesoscopic systems, it changes
the sign at high biases for any asymmetry of the cavity except 
the ideally symmetric case.
For the symmetric cavity we found a saturation of the third cumulant 
with bias at the value proportional to the temperature and the number
of channels in the leads.
We explain the change of sign for the third cumulant as a feature of 
skewness of the distribution function of transmission eigenvalues. 
This is confirmed by the analysis of the temperature dependent third cumulant 
for an arbitrary mesoscopic conductor in section \ref{sec-3arb}.
In the Appendix we show the correspondence between the CGF obtained in 
the present paper and the RMT results in the zero temperature limit.

\section{Cumulant generating function}
\label{sec-CGF}
\begin{figure}
\epsfxsize=8.0cm
\epsfbox{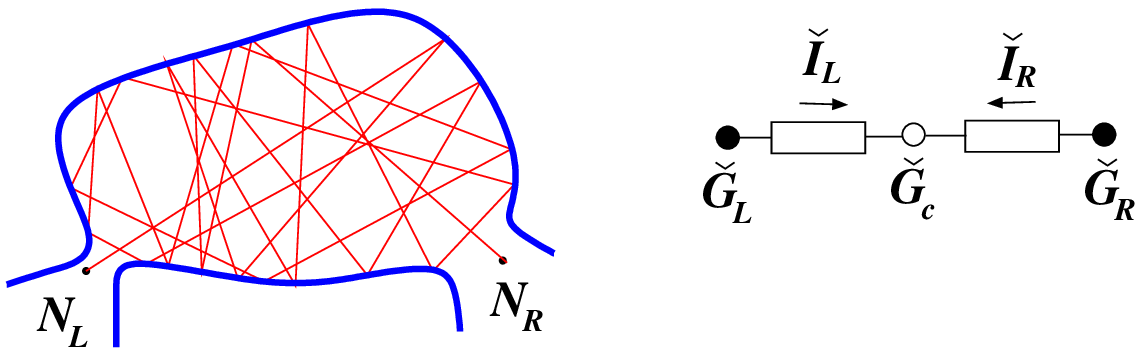}
\caption{Chaotic cavity with two asymmetric leads supporting 
$N_L$ and $N_R$ quantum channels (left) 
and its circuit theory representation (right) with matrix Green functions
$\check G_L$, $\check G_c$, and $\check G_R$ assigned to each terminal and
to a central node. The matrix currents $\check{I}_L$ and $\check{I}_R$ obey
a Kirchoff law in a central node.}
\label{fig-cavity} \end{figure}

A chaotic cavity is a conducting island of irregular shape connected to 
external leads by two quantum point contacts (figure \ref{fig-cavity}).
The openings are assumed to be small compared to the size of the cavity.
The leads are connected to electron reservoirs kept in local thermal 
equilibrium with chemical potentials $\mu_L$ and $\mu_R$.
The difference in chemical potentials $\mu_L-\mu_R=eV$ gives rise to a
current $I$ flowing through the cavity.
For simplicity, the quantum point contacts are assumed to be completely
transparent, i.e., their widths are adjusted to conductance plateaus 
\cite{oberholzer01}.
We disregard possible Coulomb blockade effects, which is justified if 
the conductances of the leads $G_L=G_0 N_L$,
$G_R=G_0 N_R$ are assumed to be much larger than the conductance quantum
$G_0=e^2/\pi\hbar$. This means a multichannel transport regime and 
negligible weak localization effects.
The inelastic relaxation inside the cavity is disregarded, so the energy
of each electron entering the cavity is conserved.
Since we are interested only in low-frequency limit of current fluctuations,
the screening effects in the cavity are neglected. It is also assumed that 
the measurement circuit has negligible impedance in order to avoid 
the environmental effects on the statistics of current fluctuations 
\cite{beenakker03}.

We apply a circuit theory of FCS based on equations for the Keldysh-Green
functions \cite{nazarov99-1}.
Its main idea is to represent a mesoscopic system by a set of discrete 
elements: terminals, nodes and connectors. 
In contrast to conventional classical circuits,
in which the nodes are described by a scalar electric potential,
here the nodes are characterized by matrix Green functions.
The matrix currents flowing between the nodes and obeying matrix analogs 
of Kirchoff rules contain all current correlators
\cite{nazarov99-1,nazarov99-2,belzig01,bagrets02,belzig-book}.

The circuit theory representation of a chaotic cavity is depicted
in figure \ref{fig-cavity} (right panel). 
It consists of two terminals (the left and right reservoirs),
a central node (the cavity) and two connectors 
(the left and right point contacts).
The terminals and the node are described by 2$\times$2 matrix Green functions 
in Keldysh space: $\check G_L$, $\check G_R$, and $\check G_c$, respectively.
The terminal matrices are assumed to be known and they are given by
\begin{equation} \label{term-matrix}
\check G_L = \rme^{\rmi \lambda\check\tau_z/2} \check G_L^0
\rme^{-\rmi \lambda\check\tau_z/2}, \quad 
\check G_R = \check G_R^0,
\end{equation}
\begin{equation} \label{term-matrix-zero}
\check G_k^0 = \left(
\begin{array}{cc}
 1-2f_k & -2 f_k \\
 -2(1-f_k) & 2f_k-1
\end{array} \right), \quad k=L,R,
\end{equation}
where the counting rotation is applied to the left terminal, 
$\tau_z$ is the third Pauli matrix and the energy dependent Fermi 
function $f_k(\varepsilon)=\{1+ \exp[(\varepsilon-\mu_k)/k T_k] \}^{-1}$
accounts for the electrochemical potentials $\mu_k$ and the
temperatures $T_k$ in the reservoirs.
For a two-terminal conductor, it is sufficient to introduce only one counting 
field $\lambda$ by gauge transformation in the left terminal.
Since we focus only to low-frequency limit of current fluctuations,
$\lambda$ is taken as a time independent constant.

The matrix current between each connector $k$ and a central node
is given by \cite{nazarov99-2} 
\begin{equation} \label{matrcurr}
\check{I}_k =\sum_n 
\frac{T_n^k [\check G_k,\check G_c]}
{4+T_n^k (\{\check G_k, \check G_c\} - 2)}.
\end{equation}
Each connector $k$ is characterized by a set of transmission 
eigenvalues $T_n^k$ associated with $n$ quantum channels at energy 
$\varepsilon$. 
For fully transparent ballistic point contacts $T_n^k=1$ for $n\leq N_k$ and 
$T_n^k=0$ for $n > N_k$, that corresponds to having $N_L$ and $N_R$ 
open channels. In general $N_L\neq N_R$.
Thus the matrix current conservation $\check{I}_L + \check{I}_R=0$ gives
\begin{equation} \label{conserv}
N_L \, \frac{[\check G_L,\check G_c]}{2+\{\check G_L, \check G_c\}} +
N_R \, \frac{[\check G_R,\check G_c]}{2+\{\check G_R, \check G_c\}} = 0.
\end{equation}
Here, the only unknown function is the matrix $\check G_c$, 
which is isotropic within the cavity under a quantum chaotic regime.
Since the terminal matrices $\check G_k$ and the node matrix $\check G_c$ 
are traceless, they can be expanded over the Pauli matrices \cite{bagrets02}:
$\check G_k = \vec{v}_k \check{\vec{\tau}}$,
$\check G_c = \vec{v}_c \check{\vec{\tau}}$, 
where $\vec{v}_k$ and $\vec{v}_c$ are three-dimensional vectors, and
$\check{\vec{\tau}} = (\check \tau_x, \check \tau_y,\check \tau_z)$ 
is the vector of the Pauli matrices.
With the use of this parametrization the anticommutators
$\{\check G_k, \check G_c\}=2\vec{v}_k\cdot\vec{v}_c$ and 
$\{\check G_L, \check G_R\}=2\vec{v}_L\cdot\vec{v}_R$ are 
scalar functions.
Introducing the notation
$\check G_{\Sigma} \equiv p_L \check G_L + p_R \check G_R$ with
\begin{equation} 
p_k \equiv \frac{N_k}{2+\{\check G_k, \check G_c\}},
\end{equation}
we rewrite equation (\ref{conserv}) as $[\check G_{\Sigma},\check G_c] = 0$,
from which we obtain $p_L\vec{v}_L+p_R\vec{v}_R=c\,\vec{v}_c$, 
where $c$ is the constant to be found. 
Using the normalization conditions 
$\vec{v}_k^2$=$\vec{v}_c^2$=1 and denoting $N_+=(N_L+N_R)/2$,
we obtain the set of equations
\begin{eqnarray} \label{set}
&c^2 = p_L^2 + p_R^2 + p_L p_R \,\{\check G_L, \check G_R\}, \\
&2p_L = N_+ - c N_R/N_+, \label{pL} \\
&2p_R = N_+ - c N_L/N_+,
\end{eqnarray}
from which the constant $c$ can be found explicitly:
\begin{equation} \label{c}
c = N_+ (1+\sqrt{1-a})^{-1},
\end{equation}
\begin{equation} \label{a}
a = \frac{4N_L N_R}{(N_L+N_R)^2} 
\left(1-\frac{4}{2+\{\check G_L, \check G_R\}}\right).
\end{equation}

Our aim is to find the CGF $S(\lambda)$ related to the matrix current by
\cite{nazarov99-2} 
\begin{equation} \label{CGF-general}
\frac{\partial S}{\partial \lambda} = \rmi t
\int \frac{\rmd\varepsilon}{\pi\hbar} \,\Tr(\check \tau_z \check I_L),
\end{equation}
where $t$ is the observation time, which is assumed to be much longer than
both the correlation time of current fluctuations and the characteristic time 
$e/I$ with $I$ the time-averaged current. The latter condition ensures that
the number of electrons transfered during the time $t$ is much greater than 1.
Substituting $\check I_L=p_L\,[\check G_L,\check G_c]$ and doing matrix 
transformations we obtain 
\begin{eqnarray} \label{CGF-A}
\frac{\partial S}{\partial \lambda} = t 
\int \frac{\rmd\varepsilon}{\pi\hbar} \, {\cal A},
\end{eqnarray}
with
\begin{eqnarray}
{\cal A} = \frac{p_L p_R}{c} \frac{\partial}{\partial
  \lambda}\{\check G_L, \check G_R\}. \label{A1}
\end{eqnarray}
To find the CGF from equation (\ref{CGF-A}) explicitly, we have to integrate 
${\cal A}$ over $\lambda$. To this end all the variables in equation 
(\ref{A1}) can be expressed in terms of the anticommutator 
${\cal G}\equiv\{\check G_L,\check G_R\}$ by means of equations 
(\ref{set})--(\ref{a}). We get
\begin{equation} \label{A-RMT}
{\cal A} = N_+ \frac{1-\sqrt{1-a}}
{{\cal G}-2}\frac{\partial{\cal G}}{\partial \lambda},
\end{equation}
where the function $a(\lambda)$ is determined by equation (\ref{a}).
Then we use the parametriza\-tion $a=4x(1-x)$, that leads to
\begin{equation} \label{Ax}
{\cal A} = 2 N_+ \frac{(1-2x)}{x [1- x(1-x)F^{-1}]} 
\frac{\partial x}{\partial \lambda},
\end{equation}
where
\begin{equation} \label{F}
F=\frac{N_L N_R}{(N_L+N_R)^2}.
\end{equation}
is the constant which will appear below as the Fano factor.
Integrating equation (\ref{CGF-A}) over $\lambda$ by making use of 
equation (\ref{Ax}), we obtain the final formula for the CGF
\begin{eqnarray} \label{CGF}
S(\lambda) = t \,(N_L+N_R) \int \frac{\rmd\varepsilon}{\pi\hbar} \,
S_{\varepsilon}(\lambda), \\ \label{CGF_eps}
S_{\varepsilon}(\lambda) = 
\alpha_L \ln \left( \frac{\alpha_L}{1-\alpha_R/x} \right)
+ \alpha_R \ln \left( \frac{\alpha_R}{1-\alpha_L/x} \right),
\end{eqnarray}
with
\begin{eqnarray}
x(\lambda) = [1+\sqrt{1-a(\lambda)}]/2, \\
\alpha_k = \frac{N_k}{N_L+N_R}, \quad k=L,R. \label{alpha}
\end{eqnarray}
It is important that the analytical result for the CGF
is expressed through the anticommutator 
$\{\check G_L, \check G_R\}$ of arbitrary (traceless) $2\times 2$ matrix 
Green functions at the terminals. 
Therefore the obtained formula is valid for different types of contacts.
\footnote[1]{
After submission of the first version of this paper to the e-archive 
[{\it Preprint} cond-mat/0403388], 
Vanevic and Belzig have proved \cite{vanevic-belzig} 
that formula (\ref{CGF}) for the CGF is independent of the matrix structure 
and, in particular, it is valid for $4\times 4$ matrices
in Keldysh-Nambu space for superconductor contacts.}

For normal metals with the help of equation (\ref{term-matrix-zero}), 
the anticommutator $\cal G$ reads
\begin{equation}
{\cal G} = 2 + 4 
[f_L(1-f_R)(\rme^{\rmi\lambda}-1)+f_R(1-f_L)(\rme^{-\rmi\lambda}-1)].
\end{equation}
It is worth noticing that our results are in agreement with the RMT theory. 
As we show in the Appendix, the distribution of transmission eigenvalues 
$\rho({\cal T})$ obtained from equation (\ref{CGF}) in the zero temperature
limit gives the RMT result.
We have also verified the correspondence of our results with the stochastic
path integral approach. The saddle-point solutions of the action in 
\cite{pilgram02} can be found not only in the symmetric limit, but quite 
generally for an arbitrary asymmetric cavity, and we have checked that the result
is identical to equation (\ref{CGF}). 

The semiclassical electron distribution function $f_c$ 
inside the cavity can be obtained from the matrix Green function 
in the central node
\begin{equation}
\check G_c=\alpha_L \check G_L + \alpha_R \check G_R + [x(\lambda)-1]
(\check G_L + \check G_R).
\end{equation}
In the limit $\lambda\to 0$ we obtain
$\check G_c=\alpha_L \check G_L^0 + \alpha_R \check G_R^0$, which with the 
help of equation (\ref{term-matrix-zero}) gives the distribution function
in the cavity $f_c=\alpha_L f_L + \alpha_R f_R$ derived earlier by
other methods \cite{blanter00}.

In the symmetric limit $N_L$=$N_R$=$N$, $\alpha_L$=$\alpha_R$=1/2, 
the CGF in equation (\ref{CGF}) is simplified to
\begin{equation} \label{CGF-sym}
S_{\varepsilon}^{\rm sym}(\lambda) = \ln[(2+\sqrt{2+{\cal G}})/4],
\end{equation}
which corresponds to the result derived by Belzig \cite{belzig-book}
with the normalization $S(0)=0$.

It is instructive to compare the symmetric and asymmetric cases in the
zero temperature limit.
We present the results for the characteristic function $\chi\equiv\rme^S$:
\begin{eqnarray} \label{char-sym}
\chi^{\rm sym}(\lambda) = 
\left(\frac{1}{2}+\frac{1}{2}\rme^{\rmi\lambda/2}\right)^{2MN}, \\ 
\chi(\lambda) = 
\left(\alpha_L+\frac{\alpha_L\alpha_R}{x(\lambda)-\alpha_R}\right)^{M N_L} 
\left(\alpha_R+\frac{\alpha_L\alpha_R}{x(\lambda)-\alpha_L}\right)^{M N_R}.
\label{char}
\end{eqnarray}
Here, $M\equiv eVt/\pi\hbar$ is the number of attempts per channel, 
which is the maximal number of electrons that can be transferred under
the restriction imposed by the exclusion principle for fermions \cite{LLL}.
Equation (\ref{char-sym}) has been obtained by Blanter \etal from 
the RMT theory \cite{blanter01} 
and by Pilgram \etal by the stochastic path integral method \cite{pilgram03}.
Equation (\ref{char}) is an extension to the case of cavities 
with non-equivalent leads, in which the relative numbers of channels 
$\alpha_k$ are defined by equation (\ref{alpha}).
The particular form of $\chi(\lambda)$ can be attributed to interplay of 
binomial statistics of the two point contacts in series (cavity openings) 
with inter-channel mixing inside the cavity. Therefore 
the logarithmic functions in equations (\ref{CGF}) and (\ref{CGF-sym}) 
come from the binomial statistics of two quantum point contacts \cite{LLL}.
The function $x$ in equations (\ref{CGF}) and (\ref{char}) 
and the square root in equation (\ref{CGF-sym}) account for the inter-channel 
mixing inside the cavity \cite{belzig-book}.

It is interesting to analyze formula (\ref{CGF}) in the strongly asymmetric 
limit when one opening supports many more channels than the other.
Let us suppose $N_R < N_L$.
Denoting the parameter of asymmetry $\eta=N_R/N_L\ll 1$, and expanding 
the CGF in series of $\eta$, we get 
\begin{eqnarray} \label{CGF-asym}
S_{\varepsilon}(\lambda) = 
(\ln \xi) \,\eta + (\xi^{-1}-1-\ln \xi)\, \eta^2 + \Or(\eta^3), \\ 
\xi(\lambda) = [2+{\cal G}(\lambda)]/4. \nonumber
\end{eqnarray}
In the zero temperature limit $\xi\to \rme^{\rmi\lambda}$, this simplifies to
\begin{eqnarray} 
S_{\varepsilon}(\lambda) &= \rmi\lambda \,\eta + 
(\rme^{-\rmi\lambda}-1-i\lambda)\,\eta^2 + \Or(\eta^3).
\end{eqnarray}
We observe that the leading-order term $\sim \eta$ is linear in $\lambda$.
This means that the charge transport in the leading-order is just the 
stationary current with no noise. The current is determined by the opening
with the smallest number of channels $I\approx G_0 V N_R$.
It does not mean, however, that there are no current fluctuations at all.
The second cumulant (noise power) and all higher cumulants
are determined by the next order term $(\rme^{-\rmi\lambda}-1)\,\eta^2$,
which depends on both openings.
For the noise power, for instance, we obtain $P_I\approx 2eI\,(N_R/N_L)$.
At finite temperatures the leading-order term $\sim \eta$ affects 
all the cumulants of current fluctuations [see equation (\ref{CGF-asym})].

\section{Electron counting statistics}
\label{sec-count}
Having found the CGF $S(\lambda)$, we obtain a complete information about 
the transfer of charges with time through a chaotic cavity.
The full counting statistics is determined by the distribution function $P(Q)$ 
giving the probability of transferring $Q$ charges through a cavity 
during the observation time $t$.
The distribution function is related to the CGF by Fourier transform:
\begin{equation} \label{P}
P(Q) = \int_{-\pi}^{\pi} \frac{\rmd\lambda}{2\pi} 
\rme^{-\rmi\lambda Q + S(\lambda)}.
\end{equation}
This integral can be evaluated in the stationary phase approximation.
In the low-temperature limit the stationary point is determined by
\begin{equation} \label{saddle}
\left. (N_L+N_R)\, M \,\frac{\partial S_{\varepsilon}}{\partial \lambda} 
\right|_{\lambda_s} = \rmi Q.
\end{equation}
The large number of channels $N_{L,R}\gg 1$ and long observation time $M\gg 1$ 
provide a necessary condition for the observation of a large number
of transmitted particles and accurate estimate of the integral 
in the stationary point.
By using the CGF (\ref{CGF_eps}) we arrive at the pure imaginary value 
for the stationary point
\begin{equation} \label{lambda_s}
\lambda_s = \rmi \, \ln 
\left[\left(\frac{MN_L}{Q} - 1\right)\left(\frac{MN_R}{Q} - 1\right)\right].
\end{equation}
The maximal number of electrons that can be transferred through each contact 
separately is given by the product $MN_k$ (the number of attempts per channel 
multiplied by the number of channels) with $k=L,R$ . 
Therefore the charge $Q$ transmitted through a cavity is restricted by 
the limits: $0<Q<Q_{max}$, where $Q_{max}\equiv MN_R$ 
and $N_R$ is the number of channels in the narrowest contact.
We notice that in this interval of allowed values of $Q$ the expression
in the logarithm (\ref{lambda_s}) is strictly positive and the stationary point
does exist.
Evaluation of the integral (\ref{P}) in the stationary point (\ref{lambda_s})
leads to the result
\begin{equation} \label{lnP}
\ln P(Q) =  - 2Q\ln Q - \sum_k (Q_k-Q)\ln (Q_k-Q),
\end{equation}
where $Q_k\equiv M N_k$, $k=L,R$.
It is interesting to see that the probability function is factorized into 
separate contributions for each contact:
$\ln P(Q)= \sum_{k=L,R} L_k$ with $L_k=-Q\ln Q-(Q_k-Q)\ln (Q_k-Q)$.

\begin{figure}
\epsfxsize=8.0cm
\epsfbox{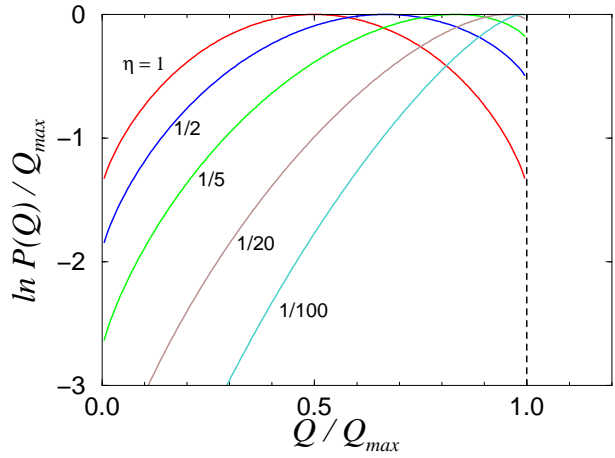}
\caption{Counting statistics of charge $Q$ transmitted trough an asymmetric 
cavity for several parameters of the asymmetry $\eta=N_R/N_L$.
The narrowest contact $R$ is kept fixed, while the contact $L$ is widened.
The charge is normalized to its maximum value $Q_{max}=(e^2Vt/\pi\hbar)N_R$ 
determined by the conductance of the narrowest opening $R$, 
and the probability is normalized according to $P(\bar{Q})=1$ 
at the point of the average transmitted charge $\bar{Q}=Q_L Q_R/(Q_L+Q_R)$.}
\label{fig-PQ} \end{figure}

The result of the evaluation of $\ln P(Q)$ using Eq.\ (\ref{lnP}) is shown in
Fig.\ \ref{fig-PQ} for different asymmetries of the cavity $\eta=N_R/N_L$.
For a symmetric cavity ($\eta$=1) the distribution function is symmetric around
the average value $Q/Q_{max}=0.5$ and it is in agreement with the known result:
$\ln P_{\rm sym}(Q)/Q_{max} = -2\, [\ln 2 + {\cal Q}\ln {\cal Q}
+(1-{\cal Q})\ln(1-{\cal Q})]$
with ${\cal Q}\equiv Q/Q_{max}$ (see, e.g., Ref.\ \cite{sukhorukov-ob05}).
When $\eta$ is decreased from 1 to $1/100$ the cavity becomes more and more 
asymmetric, the maximum $P(Q)$ is shifted from 0.5 to 1 
(see Fig.\ \ref{fig-PQ}).
This evolution correponds to the widening of one of the contacts, while 
the other (the narrowest one) is kept fixed. 
In the strongly asymmetric limit when one opening supports many more channels
than the other ($\eta\ll 1$), the distribution function becomes highly 
asymmetric with a long tail at $Q<\bar{Q}$ and a sharp drop to zero at
$Q>\bar{Q}$. The maximum of $P(Q)$ approaches the largest value 
$\bar{Q}\to Q_{max}$ determined by the smallest opening. 
This is in agreement with the fact that 
the conductance of the cavity (the first-order cumulant) in this limit 
depends only on the smallest opening. 
The width of $P(Q)$ determines the second cumulant (the noise power).
In the strongly asymmetric limit the second cumulant and all higher cumulants 
are determined by the tail of $P(Q)$ at $Q<\bar{Q}$ described asymptotically 
as $\ln P(Q) \sim (Q_{max}-Q)\ln(N_R/N_L)$ anf therefore they depend on both 
openings. This clearly shows the non-Gaussian character of the fluctuations 
of the transmitted charge in asymmetric cavities.

\section{Cumulants}
\label{sec-cum}
The cumulants are found from the CGF by differentiating
over the counting field $\lambda$:
\begin{equation} \label{ck}
C_k = \left( \frac{\partial}{\rmi \partial\lambda} \right)^k S(\lambda)
|_{\lambda=0}.
\end{equation}
The first few cumulants characterize the features of the distribution 
$P(Q)$: 
the first cumulant gives the mean, the second, the dispersion, and 
the third, the skewness of the distribution \cite{levitov01}.

\subsection{Mean current and noise}

Although the first two cumulants for the chaotic cavity are already known,
it is instructive to see how they emerge in our approach in order to compare
with higher cumulants.
The first two cumulants of current fluctuations determine the mean current 
and noise:
\begin{equation} \label{C12}
I = \frac{e}{t}\, C_1, \qquad P_I = \frac{2 e^2}{t}\, C_2,
\end{equation}
From equations (\ref{CGF}) and (\ref{ck}) we obtain 
\begin{equation} 
I = e \frac{N_L N_R}{N_L+N_R} \int \frac{\rmd\varepsilon}{\pi\hbar} \,
(f_L-f_R),
\end{equation}
\begin{equation}
\fl P_I = 2 e^2 \frac{N_L N_R}{N_L+N_R} 
\int \frac{\rmd\varepsilon}{\pi\hbar} \, \left\{ f_L(1-f_L)+f_R(1-f_R)
+ \frac{N_L N_R}{(N_L+N_R)^2} (f_L-f_R)^2 \right\}.
\end{equation}
For further calculations we assume the same temperature for all the system,
$T_L=T_R=T$, and ${\rm max}(kT,eV)\ll \mu_k$, $k=L,R$. 
We get
\begin{eqnarray} \label{IGV}
I = GV, \\
P_I = 4 kT G [1 + F \,(u \coth u - 1)], \label{PI}
\end{eqnarray}
where $G=G_L G_R/(G_L+G_R)$ is the conductance of the cavity, 
$u\equiv eV/(2kT)$, and $F$ is the Fano factor given by equation (\ref{F}).
Equation (\ref{PI}) has also been obtained semiclassically
\cite{blanter00,oberholzer01}.
It is seen that the current $I$ is linear in bias for arbitrary temperature
and therefore contains little information about the transport.
In contrast, the behavior of the noise power differs in the two limits:
\begin{equation} \label{PI-as}
P_I = \left\{
\begin{array}{lc}
4 kT G [1+ (F/12) (eV/kT)^2], & eV\ll kT \\
2eG V F, & eV\gg kT.
\end{array} \right.
\end{equation}
At low biases (high temperatures) the second cumulant, as an even-order 
cumulant, is dominated by thermal noise $4 kT G$.
To extract information additional to that
obtained from the mean current, one should go to the shot-noise regime 
$eV\gg kT$. The measurement of the shot-noise suppression factor $P_I/2qI=F$,
gives in principle information about the cavity openings.
In these measurements a natural difficulty is the requirement to maintain 
the sample cool under high currents, and to unambiguously interpret 
the linear dependence of $P_I$ over $I$ as a shot noise and not as a thermal
noise modified by nonlinear conductance \cite{levitov01}.
In contrast, the third cumulant is free from this difficulty as will be shown
in the next section.

\subsection{Third-order cumulant}

\begin{figure}
\epsfxsize=8.0cm
\epsfbox{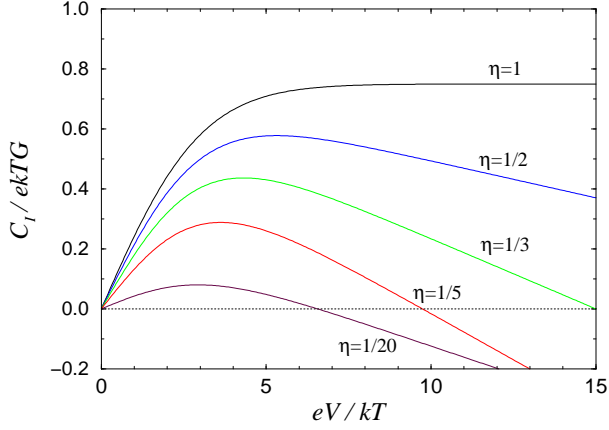}
\caption{Third cumulant of current fluctuations $C_I$ as a function of 
bias-to-temperature ratio for different degrees of the cavity asymmetry 
$\eta=N_R/N_L$.
With increasing bias, $C_I$ changes sign for any $\eta$ except $\eta$=1
for the ideally symmetric case, for which it saturates.
For negative biases $C_I(-V)=-C_I(V)$.}
\label{fig-CI-eta} \end{figure}

For the third cumulant of current fluctuations
\begin{equation} 
C_I = \frac{e^3}{t}\, C_3,
\end{equation}
from equation (\ref{CGF}) we find 
\begin{eqnarray} \label{CI-gen}
C_I = e^3 \frac{N_L N_R}{N_L+N_R} \int \frac{\rmd\varepsilon}{\pi\hbar} \, 
(f_L-f_R) \left\{ 1 - 3\gamma_1 [f_L(1-f_R) \right. \nonumber \\  
+f_R(1-f_L)] + \left. 2\gamma_2 (f_L-f_R)^2 \right\},
\end{eqnarray}
where $\gamma_1 = 1 - F$ and $\gamma_2 = 1- 2F +2F^2$.
Integrating over energy, we get
\begin{equation} \label{CI-u}
C_I = e kT G \cdot 6F(1-2F) \left[ \coth u + u 
\left(\frac{2(1-F)}{3(1-2F)} -\coth^2 u \right) \right],
\end{equation}
which is valid for arbitrary ratio $u=eV/2kT$ between the bias and temperature.
Asymptotically, 
\begin{equation} \label{CI-as}
C_I = \left\{
\begin{array}{lc}
e^2 IF, & eV\ll kT \\
e G F \,[6(1-2F)\,kT - (1-4F)\,eV], & eV\gg kT.
\end{array} \right.
\end{equation}
In the zero temperature limit equation (\ref{CI-as}) is in agreement with 
the results in \cite{blanter01,nagaev02-2}:
\begin{equation} 
C_I(kT\to 0) = - e^2 I F (1-4F)
=-e^2 I \,\frac{N_L N_R(N_L-N_R)^2}{(N_L+N_R)^4}.
\end{equation}
At high temperatures we observe that close to equilibrium $C_I$ is linear 
in bias and independent of temperature, and hence not masked 
by the Johnson-Nyquist noise. 
Furthermore, it is proportional to the Fano factor $F$ and therefore 
the measurements of $C_I/e^2 I$ at arbitrarily small biases,
give directly the factor $F$ with no need to subtract the thermal noise,
in contrast to the case of the second cumulant 
[compare with equations (\ref{PI-as})].
Another important feature is the change of sign of the third cumulant 
at high biases.
This behavior is illustrated in figure \ref{fig-CI-eta} where the third 
cumulant is shown as a function of bias-to-temperature ratio for different 
asymmetries of the cavity  $\eta=N_R/N_L$. [$\eta$ is obviously related to 
the Fano factor, $F=\eta/(1+\eta)^2$].
For the symmetric case ($\eta$=1), $C_I$ approaches the bias-independent 
constant
\begin{equation} \label{CI-sym-sat}
C_I^{\rm sym}(eV\gg kT) 
= \frac{3}{4} eG kT = \frac{3}{8} eG_0 kT N
\end{equation}
proportional to the temperature $T$ and the number of channels $N$ and 
otherwise independent of the parameters of the material.
For asymmetric cavities ($0<\eta<1$), the curves change the sign
at $eV^*/kT=6(1-2F)/(1-4F)$ which approaches 6 as $\eta\to 0$.
Equations (\ref{PI-as}) and (\ref{CI-as}) also demonstrate that with
the increase of temperature under fixed bias the third cumulant saturates, 
while the second cumulant diverges with $T$.

\begin{figure}
\epsfxsize=8.0cm
\epsfbox{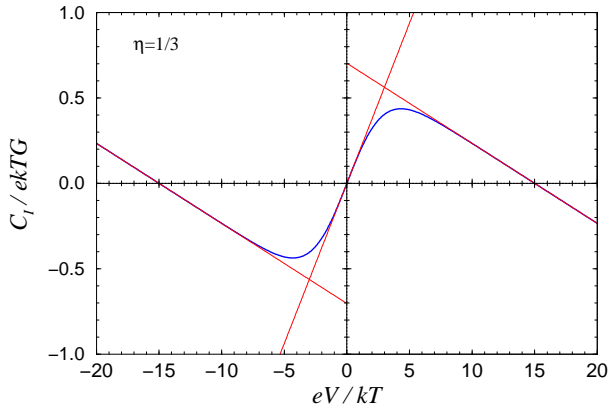}
\caption{Voltage dependence of the third cumulant $C_I$ (blue) 
for the asymmetric cavity with $\eta=1/3$.
At low and high biases $C_I$ is approximately linear (red) with positive 
and negative slopes, respectively.}
\label{fig-CI_1-3} \end{figure}

Next we focus on the voltage dependence of the third cumulant for 
the asymmetric cavity (see figure \ref{fig-CI_1-3}).
At low and high biases, $C_I$ is approximately linear, with positive and 
negative slopes, respectively. 
These linear dependences can easily be deduced from equation (\ref{CI-as}).
Note that the slopes, as well as the intersections with the axes, 
can be used in the experiment to determine 
the parameters of the cavity (the number of channels in each opening).
There is a striking similarity of our results with the behavior of $C_I$ 
for a tunnel junction in the electromagnetic environment 
modelled by two conductors in series
\cite{beenakker03}.
For the latter, the temperature dependence and the sign change are caused by 
feedback from the electromagnetic environment.
In the case of a cavity, the sign change is solely due to internal dynamics. 
This similarity is probably related to the fact that the cavity 
can also be represented by two conductors (point contacts) in series 
with inter-mode mixing in the node.

An important point is whether the temperature-dependent third cumulant 
obtained from the circuit theory [equation (\ref{CI-gen})] can also be obtained
within the semiclassical Boltzmann-Langevin picture.
The analysis of equations in the work by Nagaev \etal \cite{nagaev02-2}
shows that the same formula for $C_I$ can only be obtained if one takes into 
account the ``cascade corrections''.
[The authors present the results for $C_I$ at zero temperature only.]
At zero temperature these corrections are only significant for cumulants 
starting from the forth cumulant, but at finite temperature they are essential 
for the third cumulant too.
This means that the third cumulant we have obtained from the circuit theory
cannot be obtained from the standard Boltzmann-Langevin scheme; one has
to add the cascade correlations between the cumulants
\cite{nagaev02-2,nagaev02-1}.
In addition, it can be shown that the bias independent saturation value 
of $C_I$ for a symmetric cavity [equation (\ref{CI-sym-sat})] originates 
completely from 
the ``cascade correction'' term. The contribution of the minimal correlation
term to this value is zero.
Thus the experimental measurements of the saturation value of the third 
cumulant would give the evidence of the cascade correlations in chaotic 
cavities and the non-locality of the noise correlators.

\section{Third-order cumulant for an arbitrary mesoscopic conductor}
\label{sec-3arb}

Having found the change of sign in the third-order cumulant for chaotic 
cavities, it is interesting to ask: (i) is this behavior typical,
and (ii) what are the necessary conditions to observe the same phenomenon
in other types of mesoscopic conductors? 
Here we focus on the effects caused only by internal dynamics due to elastic
scattering; the environmental effects are neglected.

To answer these questions, we consider an arbitrary two-terminal mesoscopic 
conductor characterized by the distribution of transmission eigenvalues 
$\rho({\cal T})$. 
The CGF for this case can be found with the help of the formula obtained 
by Levitov \etal \cite{LLL}.
By expanding $S(\lambda)$ [equation (\ref{CGF-levitov}) in the Appendix] 
in Taylor series up to $\Or(\lambda^3)$, 
we obtain the third cumulant in the form \footnote[2]{
The first and the second cumulant of current fluctuations for arbitrary 
conductor are given by equations (\ref{IGV}) and (\ref{PI}), respectively, 
in which $G$ and $F$ are determined by $\rho({\cal T})$.}

\begin{equation} \label{CI-mes}
\fl C_I = e G \int \rmd\varepsilon \, 
(f_L-f_R) \left\{ 1 - 3\gamma_1 [f_L(1-f_R)
+f_R(1-f_L)] + 2\gamma_2 (f_L-f_R)^2 \right\},
\end{equation}
where $G=G_0\,\int_0^1\rho({\cal T}){\cal T} d{\cal T}$ 
is the Landauer conductance and the other constants are
\begin{equation} \label{gamma}
\gamma_1=\sigma\,\int_0^1\rho({\cal T}) {\cal T}^2 \rmd{\cal T}, \quad
\gamma_2=\sigma\,\int_0^1\rho({\cal T}) {\cal T}^3 \rmd{\cal T},
\end{equation}
with $\sigma\equiv G_0/G$.
Integrating over energy under the assumption ${\rm max}(kT,eV)\ll \mu_k$, 
$k=L,R$, we obtain
\begin{equation} \label{CI-mes-u}
C_I = e kT G 
\left\{
\beta \,[\coth u + u (1-\coth^2 u)] + 2 F_3 u \right\}
\end{equation}
as a function of the bias-to-temperature ratio $u=eV/(2kT)$.
Asymptotically, 
\begin{equation} \label{CI-mes-as}
C_I = \left\{
\begin{array}{lc}
e^2 IF, & eV\ll kT \\
e G \,[(\beta\,kT + F_3\,eV], & eV\gg kT,
\end{array} \right.
\end{equation}
where 
\begin{eqnarray} 
F&=&1-\gamma_1 
= \sigma\,\int_0^1\rho {\cal T}(1-{\cal T}) \rmd{\cal T}, \nonumber \\
\beta&=&6(\gamma_1-\gamma_2) 
= 6\sigma\,\int_0^1\rho {\cal T}^2(1-{\cal T}) \rmd{\cal T}, \label{F3} \\
F_3&=&1-3\gamma_1+2\gamma_2 
= \sigma\,\int_0^1\rho {\cal T}(1-{\cal T})(1-2{\cal T})\rmd{\cal T}. \nonumber
\end{eqnarray}
It is seen that at low biases the third-order cumulant is given by 
the universal relation $C_I=e^2 IF$ for any distribution of transmission 
eigenvalues $\rho({\cal T})$.
This means that, at low biases, $C_I$ for any conductor is: 
(i) independent of temperature, (ii) linear in voltage with a positive slope, 
and (iii) proportional to the Fano factor $F$.
At high biases, the voltage dependence of $C_I$ is also linear, with the slope
determined by the constant $F_3$.
The slopes in the low-bias and high-bias regimes are different except in the
unique case of a tunnel junction, for which all the tunneling probabilities 
are small, ${\cal T}\ll 1$, and $F\approx F_3\approx 1$, $\beta\approx 0$.
The latter is in agreement with the prediction by Levitov and Reznikov
\cite{levitov01}.

The slope of $C_I$ can be negative, that means $C_I(V)$ changes sign, 
if and only if $F_3<0$. 
Let us focus our analysis on the sign of the constant $F_3$.
We observe that the function 
$\psi({\cal T})\equiv{\cal T}(1-{\cal T})(1-2{\cal T})$ obeys 
$\psi(1-{\cal T})=-\psi({\cal T})$. Hence we can write

\begin{equation} 
F_3 = \sigma\,\int_0^{1/2}[\rho({\cal T})-\rho(1-{\cal T})] 
\,\psi({\cal T}) \rmd{\cal T},
\end{equation}
where the range of integration is reduced to $0<T<1/2$.
It is clear that the sign of this integral depends on the function
in the square brackets only, since $\psi({\cal T})>0$.
Denoting $\Delta\rho({\cal T})\equiv \rho({\cal T})-\rho(1-{\cal T})$,
we observe that $\Delta\rho({\cal T})$ is related to the asymmetry
or skewness of the distribution $\rho({\cal T})$ around the midpoint 
${\cal T}$=1/2. 
If the backscattering dominates, we can expect $\Delta\rho({\cal T})>0$ and 
$F_3>0$. If the forward transmission dominates, one can expect 
$\Delta\rho({\cal T})<0$ and $F_3<0$.
This effect is much better pronounced in the case of a one-channel conductor
\cite{levitov01},
for which $\rho({\cal T})=\delta({\cal T}-{\cal T}_1)$ and the sign of $F_3$ 
depends on the condition whether ${\cal T}_1$ is greater or smaller than 1/2.

In principle, one can calculate the coefficient $F_3$ for any mesoscopic
conductor, for which the distribution $\rho({\cal T})$ is known
(for different types of the distributions see, e.g., the reviews 
\cite{blanter-buttiker00,beenakker97}).
We just remark that for diffusive metals the distribution function is bimodal
$\rho({\cal T})\sim 1/({\cal T}\sqrt{1-{\cal T}})$ with clear asymmetry
in favor of backscattering.
One can verify that $\Delta\rho>0$ and therefore 
$C_I$ does not change sign.
Indeed, direct calculation from equations (\ref{gamma}) and (\ref{F3}) gives,
for diffusive metals, $\gamma_1$=2/3, $\gamma_2$=8/15, $F$=1/3, $\beta$=4/5,
$F_3$=1/15.
This is in agreement with the studies by Nagaev \cite{nagaev02-1} 
(semiclassical approach) and Gutman and Gefen \cite{gutman03}
(nonlinear $\sigma$-model),
who obtained the temperature dependent third cumulant for this case.

For a double-barrier tunnel junction the distribution of eigenvalues is also 
bimodal: 
$\rho({\cal T})\sim 1/\sqrt{{\cal T}^3({\cal T}_{\rm max}-{\cal T})}$ with
${\cal T}_{\rm min}<{\cal T}<{\cal T}_{\rm max}$ \cite{dejong96}.
Again, as in the case of diffusive metals, there is a skewness of 
$\rho({\cal T})$ in favor of backscattering because of two reasons: 
the shape of the distribution and the gap at high transmissions close to 
${\cal T}_{\rm max}$ (the gap at small transmissions is much smaller
\cite{dejong96}).
The coefficients for this case can be found: $\gamma_1=1-F$, 
$\gamma_2=(3/2)(1-F)^2$, $\beta=-3(1-F)(1-3F)$, $F_3=1-3F+3F^2$,
where the Fano factor $F=(g_L^2+g_R^2)/(g_L+g_R)^2$ is determined by
the conductances $g_L$ and $g_R$ of the barriers.
The coefficient $F_3$ in the zero-temperature limit for this case has been 
derived by de Jong \cite{dejong96}.
It is seen that $F_3>0$ for any $F$, including the case of a symmetric 
junction for which the Fano factor takes its maximal value $F$=1/2.
Therefore $C_I$ for double-barrier tunnel junctions is always positive.

For chaotic cavities, the function $\rho({\cal T})$ is bimodal again
[see equation (\ref{rho}) in the Appendix]; however, 
in contrast to the previous cases, it is asymmetric in favor of the forward
scattering due to a gap at low transmissions $0<{\cal T}<{\cal T}_c$.
Therefore for asymmetric cavities $\Delta\rho<0$.
The gap disappears only when the cavity is ideally symmetric.
In this case $\rho({\cal T})\sim 1/\sqrt{{\cal T}(1-{\cal T})}$ and
$\Delta\rho$=0. 
Indeed, from equation (\ref{CI-as}) follows that $F_3=-(1-4F)\leq 0$,
which is equal to zero only when $F$=1/4 for a symmetric cavity.
Therefore the negative slope of the voltage-dependent third cumulant
in asymmetric chaotic cavities is caused by the skewness in the 
transmissions when forward scattering prevails over backward scattering.
The voltage independent asymptotics occurs only when the function 
$\rho({\cal T})$ is symmetric with respect to the midpoint ${\cal T}$=1/2.

Summarizing these examples, we conclude that the third-order cumulant
of current fluctuations is an important quantity characterizing the
asymmetry of the distribution of transmission eigenvalues $\rho({\cal T})$ 
with respect to forward and backward scattering. 
While the average value is described by the conductance $G$, and
the ``bimodality'' (dispersion) is determined by the Fano factor $F$,
the skewness is captured by the factor $F_3$, in particular by its sign.
Therefore, the measurements of the third-order cumulant would give information
additional to that obtained from the first- and the second-order cumulants.

\section{Summary}
\label{sec-sum}

By using the extended Keldysh-Green function technique of the circuit 
theory, we have derived the analytical formula for the cumulant generating 
function of current fluctuations in a chaotic cavity with asymmetric leads 
at an arbitrary bias-to-temperature ratio.
The CGF contains a complete information about all-order cumulants, and 
it is expressed in terms of the anticommutator of the matrix Green 
functions at the leads, and therefore can be applied to different types of 
the contacts (see footnote\dag).

We emphasize the importance of knowledge of the CGF for any stochastic problem:
it gives not only the cumulant of any order, but also the probability 
distribution of a certain event including ``rare'' events (extreme-value 
statistics), which cannot be obtained from the knowledge 
of the first few cumulants.

For metal contacts with Fermi-Dirac filling factors and fully transparent 
leads, going beyond the previous studies at zero temperature, we have found 
that the third cumulant of current noise $C_I$ changes sign at high biases 
for any asymmetric cavity except in the case when the number of quantum channels 
in the two leads are equal (symmetric cavity).
This effect is attributed to the skewness of the distribution of transmission 
eigenvalues with respect to forward/backward scattering.
For the asymmetric case, $C_I$ is linear in bias at both low and high voltages 
with positive and negative slopes, respectively. 
For the symmetric cavity, $C_I$ is linear at low voltages and it is 
constant at high voltages with the value proportional to the temperature 
and the number of quantum channels. This value was found to be related to
the cascade correlations between the cumulants.
At low biases (or, equivalently, at high temperatures), 
the third cumulant is proportional to the Fano factor $F$ and independent
of temperature. This fact can be used in the experiment to measure the Fano
factor $F$ from the ratio $C_I/e^2I$ at arbitrarily small biases.
In contrast to the case of tunnel junctions, for which $C_I$ may change sign 
due to effect of the environment, in asymmetric cavities $C_I$ changes sign
due to internal dynamics.
The correspondence with other theories: the RMT theory, the cascaded Langevin
approach, and the stochastic path integral method has been discussed.

The first successful measurements of the third cumulant in tunnel junctions
\cite{reulet03,bomze05} leads us to believe that similar measurements 
can be performed for cavities and other mesoscopic systems.

\appendix
\section{Correspondence with RMT}

Levitov's formula for the CGF provides a connection between 
the FCS and the scattering properties of a two-terminal mesoscopic conductor 
\cite{LLL}:
\begin{equation} \label{CGF-levitov}
S(\lambda) = t \int \frac{\rmd\varepsilon}{\pi\hbar} \,
\int_0^1 \rho({\cal T}) \ln\left\{1 + {\cal T} 
\left[{\cal G}(\lambda)-2\right]/4 \right\} \, d{\cal T},
\end{equation}
where $\rho({\cal T})$ is the distribution of transmission eigenvalues.
In the zero-temperature limit this formula can be inverted to find 
the function $\rho({\cal T})$ from the CGF $S(\lambda)$ by analytical 
continuation in the complex plane \cite{sukhorukov-ob05}
\begin{equation}
\rho(T) = \frac{1}{\pi T^2} \,
{\rm Im} \left( \left.
\frac{\partial{\cal S_{\varepsilon}}}{\partial\Lambda}
\right|_{\Lambda\to -1/T-i0^+}\right).
\label{rhogen}
\end{equation}
where $\Lambda\equiv e^{\rmi\lambda}-1$.
One can find $\rho({\cal T})$ for an asymmteric cavity by 
substituting Eq.\ (\ref{CGF_eps}) into (\ref{rhogen}).
The shortest way, however, is to use directly the derivative from 
(\ref{A-RMT}), one obtains
\begin{equation} \label{dHdL}
\frac{\partial{\cal S_{\varepsilon}}}{\partial\Lambda} =
\frac{1-\sqrt{1-a(\Lambda)}}
{{\cal G}(\Lambda)-2}\, \frac{\partial{\cal G}}{\partial \Lambda}
= \frac{1}{\Lambda} \left( 
1-\sqrt{\frac{1+{\cal T}_c\Lambda}{1+\Lambda}} \right),
\end{equation}
where ${\cal T}_c=1-4\alpha_L\alpha_R = (N_L-N_R)^2/(N_L+N_R)^2$.
Now applying equation (\ref{rhogen}) to (\ref{dHdL}), we arrive at:
\begin{equation} \label{rho}
\rho({\cal T}) = 
\frac{1}{{\pi\cal T}} \sqrt{\frac{{\cal T}-{\cal T}_c}{1-{\cal T}}},
\qquad {\cal T}_c<{\cal T}<1,
\end{equation}
and $\rho({\cal T})=0$ at $0<{\cal T}<{\cal T}_c$.
This coincides with the distribution obtained by Beenakker \cite{beenakker97}
from the random matrix theory based upon the circular unitary ensemble 
of chaotic cavities.

\ack
The author would like to thank Eugene Sukhorukov for stimulating discussions
and Wolfgang Belzig, Markus B\"uttiker, Sebastian Pilgram, and
Peter Samuelsson for valuable comments.
This work was supported by the Ministerio de Ciencia y Tecnolog\'{\i}a 
of Spain through the ``Ram\'on y Cajal'' program.

\Figures

\end{document}